# The Bell inequalities: identifying what is testable and what is not

Louis Sica[1,2]

[1]Institute for Quantum Studies, Chapman University, Orange, CA & Burtonsville, MD, 20866, USA

[2]Inspire Institute, Inc., Alexandria, VA 22303, USA

Email: lousica@jhu.edu

The Bell inequalities in three and four correlations are re-derived in general forms showing that three and four data sets, respectively, identically satisfy them regardless of whether they are random, deterministic, measured, predicted, or some combination of these. The Bell inequality applicable to data is thus a purely mathematical result independent of experimental test. Correlations of simultaneously cross-correlated variable pairs do not in general all have the same form, and vary with the physical system considered and its experimental configuration. It is the form of correlations of associated data sets that may be tested, and not whether they satisfy the Bell inequality. In the case of pairs of spins or photons, a third measured or predicted value requires a different experimental setup or predictive computation than is used to obtain data from pairs alone. This is due to the quantum non-commutation of spin and photon measurements when there is more than one per particle of a pair. The Wigner inequality for probabilities, with different probabilities for different variable pairs, may be obtained from the four variable Bell inequality under a simple symmetry condition. Neither the probability or correlation inequality is violated by correlations computed from quantum probabilities based on non-commutation.

## 1. Introduction

The purpose of this paper is to identify an oversight and accompanying errors in the logic of the Bell theorem. Violation of Bell inequalities by experimental data results from misunderstanding the nature and processing of the data to be used. While quantum mechanics as a whole is not understood, and therefore admits various interpretations, the Bell inequality rests on mathematical logic alone. This has been unrecognized due to Bell's derivation, but is not a matter of interpretation once pointed out.

The Bell [1] inequality was originally derived as part of a theorem in statistics. (See Appendix.) However, the same inequality is derivable as a purely algebraic result that must be identically satisfied by three (or four) mutually cross-correlated data sets consisting of +-1's, regardless of whether they are random or deterministic [2]. The Bell inequality is thus independent of the statistical assumptions that Bell used and that have been assumed to be necessary to its derivation. Violation of the Bell inequality results from its misuse based on ignorance of its purely mathematical basis, and most surprisingly, from ignoring the established quantum mechanical principle of non-commutation.

When the inequality is applied to random processes, its basis in simultaneous cross-correlations limits the correlation functional forms that may occur among the variables occurring in each random triplet or quadruplet realization. The inherent variability of possible correlation functions that simultaneously satisfy a given inequality accounts for a chief source of misunderstanding of the Bell theorem. A type of stochastic process known as wide sense stationary [3] (WSS) has been mistakenly assumed to characterize entanglement based correlations that involve non-commuting observables. That this error has not been recognized is possibly due to the stated belief that non-commutation is a purely quantum effect [4] that should not be considered in the context of possible hidden variables in quantum mechanics. In fact however, many classical processes are non-commutative and a major example of quantum non-commutation, the Pauli spin matrices, originated in a classical representation of three dimensional rotations by two dimensional matrices [5]. This, and the non-commutation of sequences of classical light polarization measurements, indicate that *non-commutation is a fact that cannot be neglected in either classical or quantum physics.* Indeed, while non-commutation is common in the classical world, the author is not aware of any books that treat it in the context of random variables until the recent monograph by Khrennikov [6]. Finally, although encountered frequently in everyday life, there is no commonly recognized English language term for non-commutation.

Bell based his theorem on the use of counterfactuals [1] which some believe to be inapplicable to quantum mechanics. However, quantum mechanics may be used to provide a probabilistic description of certain physical processes, i.e., give predictions of results of measurement operations whether performed or not (counterfactuals). Measurements may then be carried out to confirm the predicted correlations based on the probabilities used [7]. The Bell inequalities may thus be applied to either quantum counterfactuals or performable quantum measurements as will be illustrated below.

A common explanation for the violation of the Bell inequality is that due to non-locality, more than three variables (or four) are actually interacting to produce the Bell cosine correlation. Thus, the inequality is judged to be intrinsically inapplicable to the real physical situation. However, due to the extreme generality of the Bell inequality as derived below, pickup between detectors cannot alter the fact of inequality satisfaction, although alteration of the form of correlation functions would occur.

Finally, it is quite common to consider a probability version of the Bell inequality that some seem to assume is logically independent of the correlation form that Bell derived. However, since correlations may be expressed in terms of the probabilities that yield them, it is not surprising to find that the probability form follows from the Bell inequality in correlations. Thus the probability or Wigner [8] form of the Bell inequality is not logically independent of the correlational form and is satisfied by properly constructed quantum probabilities [9].

An earlier version of this paper was posted in the quant-ph archives [10].

## 2. Bell inequality for data sets

The Bell inequality for data sets will now be derived, since it is of critical importance to understanding the Bell theorem. Data sets are defined, herein, as existing if they can be written down. They may result from experimental observations, from theoretical predictions of experimental observations (counterfactuals), or from a combination of the two. The data may also be either random or deterministic. In constructing his theorem and inequality, Bell assumed three measurement readouts from two entangled particles in a Bell-correlation measurement apparatus (Fig. 1). There appear to be only two ways to obtain such data: One conceptually simple way is to add detectors (Fig. 2) beyond the two needed to observe the correlated count pairs produced by the source. Such an arrangement immediately implies different correlations among different variable pairs.

Bell specifically rejected this approach [1]. He employed a second method using quantum probabilities to predict the result of an additional measurement at an alternative detector setting for a previously measured particle. However, only one measurement on each particle is commutative. In quantum mechanics, if two operators corresponding to measurement operations commute, they have a set of eigenstates or eigenfunctions in common. If they do not commute, their eigenfunctions are from different function sets with members of either set expressible as a linear combination of those of the other (see basic quantum mechanics texts, e.g., Mandl [11]). Thus, the probabilites for predicted correlations resulting from non-commuting alternative observables at different instrument settings should be different, and they are as will be shown below. (See Ref. [7] for a suggested expermental test of this.)

The Bell inequality will now be shown to hold under far more general circumstances than is apparent from Bell's derivation as part of a statistics theorem [2]. Assume that three data sets, random or deterministic, labeled *a, b,* and *c* have been obtained. The data set items are denoted by $a_i$, $b_i$, and $c_i$ with $N$ items in each set. Each datum equals $\pm 1$. One may form the equation

$$a_i b_i - a_i c_i = a_i (b_i - c_i) \ , \tag{2.1}$$

and sum this equation over the $N$ data triplets from the data sets. After dividing by $N$, one obtains

$$\frac{\sum_i^N a_i b_i}{N} - \frac{\sum_i^N a_i c_i}{N} = \frac{\sum_i^N a_i b_i (1 - b_i c_i)}{N} \ . \tag{2.2}$$

Taking absolute values of both sides,

$$\left|\frac{\sum_i^N a_i b_i}{N} - \frac{\sum_i^N a_i c_i}{N}\right| = \left|\frac{\sum_i^N a_i b_i (1 - b_i c_i)}{N}\right| \le \frac{\sum_i^N |(1 - b_i c_i)|}{N} \;, \tag{2.3}$$

or

$$\left|\frac{\sum_i^N a_i b_i}{N} - \frac{\sum_i^N a_i c_i}{N}\right| \le 1 - \frac{\sum_i^N b_i c_i}{N} \;. \tag{2.4}$$

Inequality (2.4) has the same form as the three variable inequality derived by Bell for correlations but is expressed in a form directly applicable to laboratory data. The algebraic steps used in it's construction are the same as those applied by Bell to previously averaged correlation functions. (See Appendix.)

The author unexpectedly discovered this result some time ago [2] by asking the following question: If one performs a laboratory experiment for which the number of data items $3N$ is necessarily finite, to what extent do random fluctuations of the correlation estimates result in violation of the Bell inequality (2.4)? Surprisingly, the answer as shown above, is that the inequality is precisely satisfied. No assumption has been made other than that the data can be written down. Further, as long as the data can be tagged with labels *a, b,* and *c,* the inequality is satisfied even if nonlocl pickup exists between detectors. In that event, the form of the correlations would be effected but not whether the inequalitiy is satisfied. Indeed, in an extreme case, the data averaged correlations might not converge to identifiable limits, but the Bell inequality would still be satisfied for any three identified data sets. No experimental loophole in this conclusion is apparent. The inequality still follws if zeros are allowed among the +-1's.

If, however, the data derive from a random process, and the correlation estimates in inequality (2.4) converge to probabilistically computable correlations as $N \to \infty$, the resulting correlations designated by $C(x,y)$ must then satisfy an inequality of the same form as inequality (2.4):

$$|C(a,b) - C(a,c)| \le 1 - C(b,c) \;. \tag{2.5a}$$

where
$$C(a,b) = \lim_{N \to \infty} \sum_i \frac{a_i b_i}{N} \;,$$

and the limit is statistical. (Bell reverses the sign of the correlation on the right side of (2.5a) by considering this *b*-measurement setting to occur on the opposite side of the apparatus from the *c*-measurement. See Sec. 4.)

This is the inequality derived by Bell [1] in which detectors at the same settings on opposite sides of a (Type II down-converter) source of entangled particle pairs produce results of opposite sign (Fig. 1). In the optical case two polarizations occur. Counts of one polarization are labeled +1 and those of the orthogonal polarization are labeled -1. Using $b'$ for the alternative measurement instead of c on the right-hand-side, the inequality is written

$$\left|C_1(a,b)-C_2(a,b')\right| \leq 1 - C_3(b,b') . \qquad (2.5b)$$

Now, all *a*-measurements occur on one side of an apparatus such as in Fig. 1 and *b*-measurements on the other. (In Bell's stochastic process model, variables on opposite sides of the apparatus have opposite sign, i.e., $A(a)=-B(a)$ so as to agree with a specific example of entanglement.)

The subscripts on the correlations indicate that *they do not all necessarily have the same functional form* as follows from the lack of conditions used in deriving Eq. (2.4). (This is directly relevant to the QM case to which (2.5b) is applied, and for which the correlations are different as results from non-commutation of measurements.) It is obvious that the constant equal to 1 occurring in (2.5b) results from the fact that the value $a_i$ that multiplies $b_i$ also multiplies $b_i'$, data triplet by data triplet. However, if data are obtained from an independent run for each correlated measurement pair as is common in practice, then six data sets instead of three are used, the condition under which the inequality was derived does not hold, and the inequality will in general be violated. Strangely, the use of independent runs has become accepted experimental practice.

*It is critically important to understand that while inequality (2.4) holds generally for any three arbitrary data sets, the Bell inequalities (2.5a,b) do not hold for arbitrary correlations.* Since correlations must result from the convergence of the correlation estimates that satisfy Eq. (2.4), they must satisfy Eq. (*2.5a)* and their functional forms are mutually constrained thereby. Arbitrary correlations *not* derivable as limiting forms of correlations of three concurrently existing data sets will not necessarily satisfy inequalities (2.5a,b). If inequalities (2.5a,b) are violated by such *assumed* limiting forms, no data sets of triplets can exist that produce them.

Similar assumptions to those used to derive inequality (2.5a) from inequality (2.4) can also be used to derive a four variable Bell inequality. Assuming that there exist four data sets of size $N$ with members $a_i, a_i', b_i, b_i'$ with each datum equal to $\pm 1$, then for each group of four data items from the four data sets, one has (by inspection)

$$-2 \leq a_i(b_i+b_i') + a_i'(b_i-b_i') \leq 2 . \qquad (2.6)$$

Relation (2.6) also holds if zeros occur among the variables. Summing over $i$ from 1 to $N$ in inequality (2.6), and dividing by $N$ leads to

$$-2 \le \frac{\sum_i^N a_i b_i}{N} + \frac{\sum_i^N a_i b_i'}{N} + \frac{\sum_i^N a_i' b_i}{N} - \frac{\sum_i^N a_i' b_i'}{N} \le 2 \, . \tag{2.7a}$$

Since all experimental data sets are intrinsically finite, four data sets must satisfy inequality (2.7a), as three must satisfy inequality (2.4). Again assuming statistical convergence to limits as $N \to \infty$, a common form of Bell inequality used by experimentalists results:

$$-2 \le C_1(a,b) + C_2(a,b') + C_3(a',b) - C_4(a',b') \le 2 \, . \tag{2.7b}$$

As in the case of inequality (2.5b), the correlations may have different functional forms.

*The difficulty of applying the three variable inequality (2.4) to an entangled pair of particles in which more than two measurements are non-commutative is amplified in the case of a four variable inequality.* Note, as in the previous case, that while inequality (2.7a) must be identically satisfied by any four data sets that may be written down, inequality (2.7b) may be violated by *assumed* correlations. However, as in the three variables case, if it is violated it follows that no four data sets exist whose cross-correlations result in the assumed correlation functions.

Note that the notion of experimentally "testing" the above inequalities, in either three or four variables, involves a logical mischaracterization. Only the *form of the several correlations* that describe data from a given physical experiment may be tested, and not whether or not cross-correlations of the data satisfy the Bell inequalities. (In the laboratory counts are observed, and correlations computed from them.) Further, if variables are obtained from random realizations that yield measurements on only one pair per realization among the four correlated variables, the correlations will in general be different than if all four variable values are obtained per realization. A conceptually simple way to obtain four data outputs per realization is shown in Fig. 2.

**3. How independent data-pair correlations may violate Bell inequalities**
3.1 Bell operationally assumed correlations that are wide-sense stationary

Inequalities in three and four variables were derived from Bell's assumption of a stochastic process representation of quantum entanglement [1]. Bell represented detector readouts with a function $A(a,\lambda) = \pm 1$, where $a$ is an instrument setting and $\lambda$ denotes one or more random variables determining the resulting random values taken by $A(a,\lambda)$. In Bell's representation, there is no implication that accessing a readout at $a$ affects the probability of accessing a readout at $a'$ for a given realization. The multiple readouts of function $A(a,\lambda)$ and their associated probabilities are analogous to a set of commuting observables in QM. Howeever, for the case of non-commuting QM observables that applies here, probabilities of specific readouts at successive instrument settings are conditionally dependent on readouts at prior settings [11]. The conditional probabilites at different instrument settings have different values, whereas they would be expected to be the same if commutative states were involved.

Without stating it, Bell effectively and operationally assumed properties for the several quantum mechanical outputs with which he was concerned that correspond to a *special*, not universal, kind of random process defined as wide-sense-stationary [3] (WSS). Such a process is one in which the correlation of the readouts $A(a,\lambda)$ at any two instrument settings $a_i$ and $a_j$ is given by a function of the form $f(a_i - a_j)$ depending on the difference of coordinate settings for all setting pairs. Thus, in Bell's notation

$$f(a_i - a_j) = \int_{-\infty}^{\infty} A(a_i,\lambda) A(a_j,\lambda) \rho(\lambda) d\lambda \qquad (3.1)$$

where $\rho(\lambda)$ is a probability distribution for $\lambda$, and $a_i$ and $a_j$ are any detector settings.

Bell used the correlation functional form computed from QM for commuting measurements on a pair of entangled spins or photons that suggests a WSS process, $-\cos(a-b)$. The measurements commute because they are carred out on two different particles and it does not matter which measurement occurs first. Bell computed QM correlations at a setting $a$ for a first detector and two alternative settings $b$ and $b'$ for a second detector [1, Chap 8]. Predicted QM correlations $C(a,b)$ and $C(a,b')$ are both given by the negative cosine of detector angular differences (suggesting WSS). However, if the resulting correlation $C(b,b')$ at output settings $b$ and $b'$ is computed from the different QM probabilities that occur for each non-commuting variable and fixed value $a$ (that Bell assumed), the result is *C(a,b)C(a,b')* as will be shown below.

3.2 How misinterpretation of the Bell inequality leads to its violation

Inequalities (2.5b) and (2.7b) result from the cross-correlations of three and four data sets, respectively. A triplet or quadruplet of data values must occur in one realization of the associated random process. Whether three or four variables, each equal to +-1 are cross-correlated, determines the constant of the related inequality, 1 in the case of three variables, 2 in the case of four variables. However, if correlations are obtained from three or four variable pairs, with each pair acquired in an independent experimental run, the correlations will in general be quantitatively different from those that result from cross-correlated data triplets or quadruplets acquired in one run (as could be accomplished with the setup of Fig 2). *Different measurement scenarios for the random variables will* *in general affect both correlations between variables and their corresponding probabilities*.

Given measurement apparatus such as shown in Fig. 1, two possible scenarios are identified for acquisition of three correlations. One may measure output pairs in independent runs at settings $(a,b)$, $(a,b')$, and $(b',b)$, producing separate realizations of each variable pair. In that case, the correlation of each variable pair is given by the same function in the quantum situation under consideration, *but the conditions that lead to Bell's derivation of the inequality as well as (2.4), are violated.*

A second scenario (that specified by Bell) is to predict the three outputs at settings $a$, $b$, and $b'$ (with $b$ and $b'$ both on the right-hand side) for each random realization and calculate resulting correlations $C(a,b)$, $C(a,b')$, and $C(b',b)$ from QM probabilities for the variable pairs. Clearly the correlations and probabilities should be different from those obtained in the first scenario. Given that probabilities $P_{x,y}(a,b)$, $(x,y \in \pm 1)$ are known from QM for measurements at $(a,b)$ and $(a,b')$ given setting $a$, one can immediately compute $P(b|a)$ and $P(b'|a)$ allowing the evaluation of $P(b,b'|a) = P(b|a)P(b'|a)$. Thus, the now connected correlations of outputs at setting pairs $(a,b)$, $(a,b')$, and $(b',b)$ may be determined. Hess has pointed out [12] that similar facts and inequalities related to those of Bell have been known to mathematicians since Boole. Pure mathematics determines a third correlation when data for two correlations out of the three are specified.

3.3 WSS correlations satisfy the Bell inequality but they are not co-sinusoidal

It is instructive to consider inequalities (2.7a,b) for a finite value of $N$ in the special case of a WSS process and four data sets. The WSS properties have been assumed to represent entanglement by experimentalists and theoreticians alike, after Bell's mistaken assumption of their applicability. One may write inequality (2.7a) in the form

$$-2 \leq C(\theta_a - \theta_b) + \delta_N(a,b) + C(\theta_a - \theta_{b'}) + \delta_N(a,b') +$$
$$C(\theta_{a'} - \theta_b) + \delta_N(a',b) - C(\theta_{a'} - \theta_{b'}) - \delta_N(a',b') \leq 2 \qquad (3.2)$$

where the $C(x - y)$ functions are assumed to represent the limiting forms for the correlation estimates as $N \to \infty$. Since the inequality cannot be violated by data sets that are jointly present and cross-correlated, the $\delta_N$'s represent random differences between the infinite and finite averages of the four correlations that must sum to zero when the four variables' values are present in each realization of the experiment.

By contrast if the data are taken in four independent runs using the same instrument settings, inequality (3.2) for the same WSS process becomes

$$-2? \le C(\theta_a - \theta_b) + \delta_N(a_1, b_1) + C(\theta_a - \theta_{b'}) + \delta_N(a_2, b_2')$$
$$+C(\theta_{a'} - \theta_b) + \delta_N(a_3', b_3) - C(\theta_{a'} - \theta_{b'}) - \delta_N(a_4', b_4') \le 2? \tag{3.3}$$

where the subscripts 1…4 indicate the experimental run number used to compute the correlation statistical fluctuation, and the question marks indicate possible violation of the ±2 limits since the data are no longer cross-correlated. Eight data sets and not four are used, and the $\delta_N$'s no longer need to sum to zero, even though the limiting correlations $C(x-y)$ are given by the same function for the WSS process assumed. The error fluctuations, i.e., the $\delta_N$'s, in each of the four runs of $N$ data pairs, exist and are finite. However, assuming that the correlation estimates statistically converge to the probability averages for large data sets, the $\delta_N$'s become small as $N$ becomes large. Thus, the inequality (3.3) would be expected to be violated, but by smaller and smaller values as $N$ becomes larger.

3.4 Quantum mechanical Bell correlations cannot represent a WSS process

Bell effectively assumed [1] that the random process applicable to a triplet of polarization or spin measurement correlations is WSS, as is also widely done in the four variable case of inequality (2.7b) by those interpreting experimental data in a way that violates inequality (2.7b). When the mathematical facts leading to inequalities (2.5) or (2.7b) are considered, however, it becomes clear that *the measurement results in QM experiments do not represent a WSS process*. If they did, violation of the corresponding Bell inequality would be expected to be small, i.e., of the order of four standard deviations, rather than 102 standard deviations as has been reported [13]. Such inequality violations represent proof that the correlations of the process under consideration cannot all be co-sinusoidal and indeed they are not, as will be shown below. In the case of QM entanglement, only the measurements that commute between two particles are of this form.

**4. The Wigner inequality results from a Bell inequality if probabilities are symmetric**

A probability inequality known as the Wigner [8] inequality is intimately related to the Bell inequality constraint on correlations [9]. It relates the probabilites for pairs of +1 outcomes corresponding to Bell correlations at given instrument settings. The result is

$$P_{++}(a,b) \le P_{++}(a,c) + P_{++}(a'=c,b). \tag{4.1}$$

where the first letter in each probability $P_{++}(a,b)$ indicates the angular setting for the random variable on the left side of the Bell apparatus in Fig. 1, and the second letter indicates the angular setting for the random variable on the right side. The subscripts + and − indicate whether the variables at settings $a$ and $b$ have values of +1 or -1. Inequality. (4.1) follows from the Bell inequality

$$|C(a,b)-C(a,c)| \leq 1-C(b,c) . \quad (4.2a)$$

If setting $b$ of the right-hand-side correlation specifies instead the $a$-side setting of a Bell apparatus, the same numerical output occurs with a reversed sign under Bell's random process model where $A(b,\lambda)=-B(b,\lambda)$. Then inequality (4.2a) becomes

$$|C(a,b)-C(a,c)| \leq 1+C(b,c) . \quad (4.2b)$$

A physical process is now considered with probabilities having the same symmetry of occurence for $\pm 1's$ as occurs in QM in the case of two entangled spins. The joint probabilites resulting from an entangled spin state are:

$$P(b=\pm 1, a=\pm 1) = \frac{1}{2}\sin^2\frac{\theta_b - \theta_a}{2}; \qquad P(b=\pm 1, a=\mp 1) = \frac{1}{2}\cos^2\frac{\theta_b - \theta_a}{2} . \quad (4.3)$$

where a contracted notation indicates possible values of random variables $b$ and $a$ at settings $\theta_b$ and $\theta_a$. The probabilities conditional on $a$ are easily obtained since $P(a=\pm 1)=1/2$. The joint probabilities (4.3) thus have the following symmetry for $\pm 1$ occurrence:

$$P_{++}(a,b) = P_{--}(a,b), \ P_{+-}(a,b) = P_{-+}(a,b) . \quad (4.4)$$

The normalization condition is

$$2P_{++}(a,b) + 2P_{+-}(a,b) = 1 . \quad (4.5a)$$

so that $C(a,b)$ is given by

$$C(a,b) = 2P_{++}(a,b) - 2P_{+-}(a,b). \quad (4.5b)$$

or

$$C(a,b) = 4P_{++}(a,b) - 1 \quad (4.5c)$$

after using equation (4.5a). The results for $C(a,c)$ and $C(bc)$ are obtained by renaming the variables in equation (4.5c). The use of correlation (4.5c) in inequality (4.2b) with appropriate variables for different correlations produces:

$$4P_{++}(a,b) - 1 - (4P_{++}(a,c) - 1) \leq 1 + 4P(b,c) - 1,$$

or the Wigner inequality:

$$P_{++}(a,b) \leq P_{++}(b,c) + P_{++}(a,c). \quad (4.6)$$

As is well known, inequality (4.6) is violated when the same quantum Bell-state probability is used for all terms.

To show that quantum mechanical probabilities are consistent with a probability form of the Bell inequality it is simpler to use inequality (2.5b) since that form directly represents the physical situation considered by Bell[1]. In Chapter 8 of Bell's book, *Speakable and unspeakable in quantum mechanics*, Bell indicates that the result for the variable at $b'$ is a predicted value on the same side of the apparatus as $b$.

The computation of $C(b,b')$ in terms of $P_{++}(b,b')$ (where both *b*-variables are now on the *B*-side of the apparatus) follows if the symmetry of equations (4.4) holds for the probabilities of the observed and predicted variables:

$$C(b,b') = \sum_a [P_{++}(b,b'|a) + P_{--}(b,b'|a)]P(a) - \sum_a [P_{-+}(b,b'|a) + P_{+-}(b,b'|a)]P(a). \quad (4.7a)$$

The joint probability $P_{++}(b,b')$ computed from probabilities (4.3) is

$$P_{++}(b,b') = \tfrac{1}{2}[P_+(b|a=1)P_+[b'|a=1] + P_+(b|a=-1)P_+[b'|a=-1]],$$
$$= \tfrac{1}{2}[\sin^2\tfrac{\theta_b - \theta_a}{2}\sin^2\tfrac{\theta_{b'} - \theta_a}{2} + \cos^2\tfrac{\theta_b - \theta_a}{2}\cos^2\tfrac{\theta_{b'} - \theta_a}{2}]. \quad (4.7b)$$

A similar calculation for $P_{--}(b,b')$ yields the same result. Similarly, $P_{+-}(b,b') = P_{-+}(b,b')$ with result

$$P_{+-}(b,b') = \tfrac{1}{2}[\sin^2\tfrac{\theta_b - \theta_a}{2}\cos^2\tfrac{\theta_{b'} - \theta_a}{2} + \cos^2\tfrac{\theta_b - \theta_a}{2}\sin^2\tfrac{\theta_{b'} - \theta_a}{2}] \quad (4.7c)$$

Since the probabilities leading to $C(b,b')$ have the same symmetries as for $C(a,b)$ and $C(a,b')$, the joint probabilities for variable pairs may be put in the form of correlation (4.5c) to be used below in the computation of $C(b,b')$. However, although the specified symmetries are the same, the probabilities are very different from probabilities (4.3) that lead to the Bell correlation for the first measurement pair.

## 5. Quantum counterfactual probabilities and correlations satisfy the Bell inequality
5.1 Satisfying the Bell inequality

Inequality (4.6) is violated by quantum probabilities upon assuming that all corresponding correlations have the same form as those for the two commuting measurements. This occurs because the correlation on the right-hand-side of inequality (2.5a) is constrained by the left-hand correlations $C(a,b)$ and $C(a,c)$ whose existence requires data that determines the right-hand side. The correlation $C(b,c)$ thereby determined cannot have the same form as the previous correlations if the latter have the Bell cosine form. Repeating inequality (2.5b):

$$|C(a,b) - C(a,b')| \le 1 - C(b,b'). \quad (5.1a)$$

Using the probabilities for $P_{++}(b,b')$ and $P_{+-}(b,b')$ given above in equations (4.7b) and (4.7c) the resulting correlation $C(b,b')$ may be computed as [7]:

$$C(b,b') = 2P_{++}(b,b') - 2P_{+-}(b,b') = [\sin^2\frac{\theta_b - \theta_a}{2}\sin^2\frac{\theta_{b'} - \theta_a}{2} + \cos^2\frac{\theta_b - \theta_a}{2}\cos^2\frac{\theta_{b'} - \theta_a}{2}]$$

$$-[\sin^2\frac{\theta_b - \theta_a}{2}\cos^2\frac{\theta_{b'} - \theta_a}{2} + \cos^2\frac{\theta_b - \theta_a}{2}\sin^2\frac{\theta_{b'} - \theta_a}{2}] \quad (5.1b)$$

$$= \cos(\theta_b - \theta_a)\cos(\theta_{b'} - \theta_a).$$

The same result may also be computed by using conditional probabilites for data outputs at $b$ and $b'$ given outputs for $a$. This result could be tested experimentally as suggested in Ref [7] where an analogous result is also given in the four variable case. Using the contracted notation $\theta(b,a) \equiv \theta_b - \theta_a$, and the Bell cosine correlations, inequality (5.1a) becomes

$$|-\cos\theta(a,b) + \cos\theta(a,b')| \leq 1 - \cos\theta(b,a)\cos\theta(b',a) = 1 - \cos\theta(a,b)\cos\theta(a,b'), \quad (5.1c)$$

from which

$$\left|-2\sin\frac{\theta(a,b) + \theta(a,b')}{2}\sin\frac{\theta(a,b') - \theta(a,b)}{2}\right| \leq \sin^2\frac{\theta(a,b) + \theta(a,b')}{2} + \sin^2\frac{\theta(a,b) - \theta(a,b')}{2}, \quad (5.1d)$$

after use of appropriate trigonometric identities. One may replace the difference of correlations on the left-hand-side of inequality (5.1c) by expressions in probabilities $P_{++}$ but the same result occurs in inequality (5.1d). Since

$$(a^2 + b^2) - 2|a||b| = (|a| - |b|)^2 \geq 0,$$
$$(a^2 + b^2) \geq 2|a||b|, \quad (5.2)$$

the Bell inequality (5.1a) is satisfied.

Thus, when correlations computed from probabilities resulting from QM are used, whether expressed in correlational or probability form, inequality (5.1a) is satisfied as demanded by basic mathematics. Deductions of non-locality or non-reality, if based on Bell inequality violation, no longer follow.

5.2. Dealing with possible pickup between detectors

If measurements are made on two particles, one of the measurements occurs first, except in circumstances of infinite time precision. Assuming that $A$ is measured before $B$ or $B'$ by some time increment, any assumed pickup from other detectors by $A$ is fixed when $A$ is measured. If there is also pickup from detector $A$ to $B$ or $A$ to $B'$, three data sets are still obtained. Thus the three variable inequality holds even for the corrupted data. Observed correlation functions could then be compared with QM predictions to determine if evidence of pickup in fact exists. The Bell inequality would still be satisfied by the data sets even if the correlation estimates failed to converge to identifiable functions.

## 6. Conclusion

The principle claim with which this article is concerned is that correlations of ideal quantum mechanical data violate the Bell inequality. Since it has been shown that the same inequality holds identically for data sets as a fact of algebra without Bell's assumptions, the notion that it is testable rests on a mathematical oversight. This has resulted in misuse of an inequality that must be identically satisfied when used correctly. What may be experimentally tested is not whether the Bell inequality is satisfied when correctly used, but the form of the multiple correlation functions realized from qualitatively different measurements. If correctly computed, correlations consistent with quantum mechanics do not all have the cosine form that Bell and others have assumed based on independent count-pair measurements. The Bell inequality constants result from data triplets and quadruplets obtained per realization, and not data pairs.

Understanding of the Bell inequality follows simply in the absence of logical errors. The three and four variable inequalities are identically satisfied by cross-correlations of even ideal quantum data sets of ±1'$s$ as a fact of algebra. Their satisfaction by quantum measurements follows from a well-known quantum mechanical fact: performed measurements on spins or photons are non-commutative when more than one per particle of an entangled pair is invoked. When both facts are employed, the Bell inequality is satisfied without mysteries.

The Bell theorem may be interpreted to imply that one cannot construct a local probability model that accounts for quantum correlations without entanglement. If the logic of the theorem is flawed, however, it does not follow thereby that the converse is true. The question that immediately arises is: how much of conventional quantum mechanics is to be accounted for. Since a local algorithmic model for Bell correlations has been presented [14], the elimination of non-locality would seem to be an attainable goal. This is in agreement with the observation that the physical superposition of four waves that produce entanglement in a down-converter source no longer exists when the waves propagate to spatially separated detectors for count detection. In this case one may derive a *local* probability model resulting from the boundary conditions that exist at the source [15]. This model appears to be consistent with both quantum electro-dynamics and wave optics.

The literature relevant to Bell's theorem has grown to a large size. There are growing numbers of papers that disagree with the Bell consensus according to [16]. The author apologizes in advance to those not cited that agree with him, and to those not agreeing as well. However, this already quite long article is concerned with the logical components of this controversial topic, rather than a review of the literature. A more uniform coverage of recent contributions will have to wait until a later date.

Appendix

It is useful to derive Bell's version of his inequality by applying an explicit probability average to the left hand side of Eq. (2.2):

$$Av\left(\frac{1}{N}\sum_i^N a_i b_i - \frac{1}{N}\sum_i^N a_i b'_i\right) = \frac{1}{N}\int \sum_i^N a_i(\lambda)\left(b_i(\lambda) - b'_i(\lambda)\right)\rho(\lambda)d\lambda. \quad (A1)$$

In (A1), $\rho(\lambda)$ is a probability density for a collection of random variables $\lambda$ that determine the values of the data items $a_i, b_i$, and $b'_i$. Changing to Bell's notation for which $A(a,\lambda) = \pm 1$ at setting $a$ and $B(b,\lambda) = \pm 1$ at setting $b$, each function determined by random variable $\lambda$, Eq. (A1) becomes

$$\frac{1}{N}\int \sum_i^N a_i(\lambda)\left(b_i(\lambda) - b'_i(\lambda)\right)\rho(\lambda)d\lambda = \int A(a,\lambda)\left(B(b,\lambda) - B(b',\lambda)\right)\rho(\lambda)d\lambda, \quad (A2)$$

where the probability averages are independent of subscript $i$ since they are the same for each $i$. The three variables $A, B, B'$ have random values determined by the collection of random parameters $\lambda$, and define a stochastic process. The right side of (A2) then becomes

$$\int A(a,\lambda)\left(B(b,\lambda) - B'(b',\lambda)\right)\rho(\lambda)d\lambda =$$
$$\int A(a,\lambda)B(b,\lambda)\left(1 - B(b,\lambda)B'(b',\lambda)\right)\rho(\lambda)d\lambda. \quad (A3)$$

Taking absolute values of both sides and bringing the absolute value inside the integral on the right leads to

$$|C(a,b) - C(a,b')| \le \int |A(a,\lambda)B(b,\lambda)|\left(1 - B(b,\lambda)B'(b',\lambda)\right)\rho(\lambda)d\lambda =$$
$$\int \left(1 - B(b,\lambda)B'(b',\lambda)\right)\rho(\lambda)d\lambda = \int \left(1 - B(b,\lambda)B'(b',\lambda)\right)\rho(\lambda)d\lambda = 1 - C(bb') \quad (A4)$$

where $C(a,b)$ indicates the probability average of $A(a,\lambda)B(b,\lambda)$ etc. Relations (A4) end in Bell's inequality [1].

   Bell's notation suggests that any number of readouts may be obtained for a given realization $\lambda$ of the process. It is consistent with a WSS process as described above. The *interpretation* used thus specifies a particularly simple random process that is by no means universal. The result (2.4) proves that the same relation holds independently of Bell's stated assumptions in proving version (A4), and even if the data are deterministic.

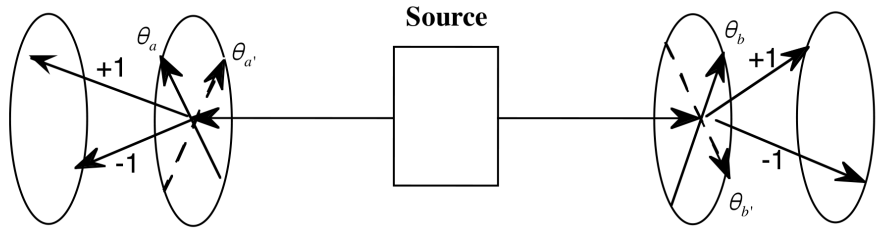

Figure 1. Schematic of Bell experiment in which a source sends two particles to two detectors having angular settings $\theta_a$ and $\theta_b$ and/or counterfactual settings $\theta_{a'}$ and $\theta_{b'}$. While one measurement operation on the *A*-side, e.g. at setting $\theta_a$, commutes with one on the *B*-side at $\theta_b$, any additional measurements at either $\theta_{a'}$ or $\theta_{b'}$ are non-commutative with prior measurements at $\theta_a$ and $\theta_b$, respectively. This figure was drawn by the author and modified in notation for use in Refs. 2, 7, 9, as well as other papers.

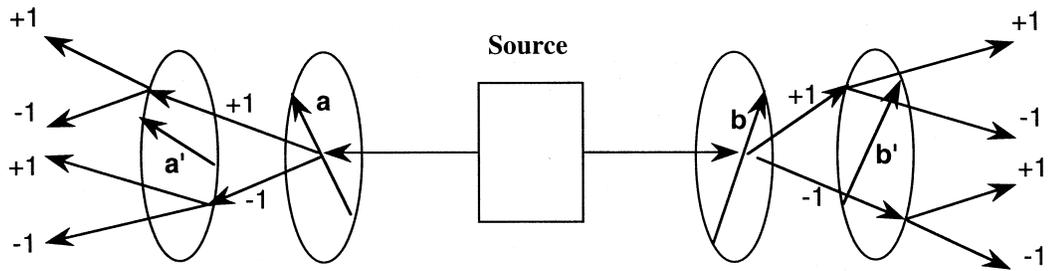

Figure 2. Schematic of a Multiple Stern–Gerlach apparatus. Arrows $a$, $a'$, $b$, $b'$ indicate the magnetic field directions encountered by pairs of particles emitted in opposite directions by the source. At each encounter with a magnetic field, the particle is deflected in one of two directions depending on whether its spin is $+1/2$ or $-1/2$. Each sequence of $\pm 1$'s corresponds to a unique spot position so that knowledge of two spin measurements is yielded retrospectively for each particle.